# Probing the Topological Surface States through Magnetoresistance and Ultrafast Charge Carrier Dynamics in $(Bi/Sb)_2Te_3$

Prince Sharma[1,2], Yogesh Kumar[1,2], V.P.S. Awana[1,2], Mahesh Kumar[1,2*]

[1]*National Physical Laboratory (CSIR), Dr. K. S. Krishnan Road, New Delhi-110012, India*
[2]*Academy of Scientific and Innovative Research, Ghaziabad U.P.-201002, India*

**Abstract**

Topological insulators with their topological protected surface states are highly promising quantum materials. In this article, the micro-flakes of single-crystalline topological insulators $Bi_2Te_3$ and $Sb_2Te_3$ are explored through physical parameter measurement at low temperatures, and thereby the charge carrier's dynamics are investigated at 5 K to study the various optical transitions related to these surface states. The magnetoresistance is experimentally investigated at temperatures of 5 K and 100 K for a field range of ± 1 Tesla. The occurrence of the weak anti-localization (WAL) effect predicts the presence of topologically protected surface states (TSS) in the systems. Further, the ultrafast femtosecond transient reflectance spectroscopy (TRUS) is performed at different temperatures, varying from a room temperature (300 K) to a low temperature of 5 K, to find the TSS-related transitions at low temperatures.

**Keywords:** Topological Insulators, Bismuth Telluride and Antimony Telluride, Magnetoresistance, Low-temperature Ultrafast Spectroscopy

# 1 Introduction

The topological insulators (TIs) are characterized through topologically protected surface states (TSS) which makes these quantum materials to be metallic in nature, but still, the bulk behaves as an insulator. The quantum materials offer numerous applications due to the presence of these TSS, which includes quantum computing[1–8], thermoelectricity[9–11], superconductivity[12–15], Majorana fermions[1,16–19], terahertz frequency generation[20–22], optoelectronics and spintronics[1,23–25]. These TSS can be theoretically probed through the density functional theory (DFT) [26–30], which were eventually confirmed through angle-resolved photoelectron spectroscopy (ARPES)[3,4,19,30–33] and scanning tunnel spectroscopy (STS)[34,35]. These techniques help in confirming the presence of TSS in a particular quantum material. Apart from these techniques, two more approaches are carried out to predict the TSS indirectly, which includes the physical property measurement of magnetoresistance at low temperature and low magnetic field[36–47], and the second is through ultrafast probing of charge carriers dynamics by femtosecond laser in TIs at low temperatures[48–50]. The V-type cusp behaviour of magnetoresistance at low temperature and magnetic field resembles the weak anti-localization (WAL) effect in TIs, further explained through the Hikami-Larkin-Nagaoka (HLN) equation[8,45,51]. The HLN modelling of magnetoresistance confirms the existence of surface states at low temperature in thin films and also in bulk flakes of single-crystal too[8,45,51].

While considering the ultrafast carrier dynamics of Tis in femtosecond and picosecond regimes, there are two ways to probe the TSS[52–54]. One is to reduce the sample thickness to a particular dimension where the bulk contribution becomes negligible, and the charge carrier's dynamics through surface states can easily be probed through TRUS [52,54,55]. While the second approach is to study the charge carrier dynamics at low temperatures where the bulk states are suppressed by the TSS even in bulk samples as measured through magnetoresistance and WAL studies. In this article, the second way is targeted where bulk micro flakes are used to study the temperature-dependent ultrafast and transport properties of TIs. These micro flakes are mechanically exfoliated using surgical blades. The X-ray diffraction (XRD), scanning electron microscopy (SEM), energy dispersive X-ray analysis spectroscopy (EDAX), and Raman spectroscopy confirm the single crystalline phase, structural, composition, and vibrational modes respectively of both bismuth telluride and antimony telluride flakes. The magnetoresistance is experimentally investigated at temperatures 5 K and 100 K in a magnetic field range up to ± 1 tesla. The occurrence of the WAL effect predicts the presence of TSS in the systems. Further, the ultrafast femtosecond transient reflectance spectroscopy (TRUS) is performed at different temperatures, which varies from a room temperature of 300 K to a low temperature of 5 K. The target is to find the TSS at these low temperatures to confirm the presence of TSS in the bulk single crystal flakes.

## 2 Experimental techniques

### 2.1 Single-crystal growth and mechanical exfoliation in micro-flakes

Single crystals of $Bi_2Te_3$ and $Sb_2Te_3$ were grown using a self-flux method. In this solid-state reaction route, high purity (4N) powders of Bi, Sb, and Te were taken in the required stoichiometric ratio and grounded thoroughly using an agate mortar pestle in a mBraun glove box filled with argon gas. The finely grounded powders were then palletized and encapsulated in a quartz tube at a vacuum of $10^{-5}$ mbar. These encapsulated samples were then placed in a PID-controlled Muffle furnace under well-optimized heat treatment, as mentioned somewhere else[20,56]. Thus, the obtained crystals were silvery shiny and easily cleavable along its growth axis using the surgical blade.

### 2.2 Structural and compositional confirmation techniques

The XRD pattern was recorded using Rigaku made table-top X-Ray Diffractometer equipped with Cu-K$_\alpha$ radiation of 1.54 Å. XRD pattern was recorded for mechanically cleaved crystal flake. At the same time, the compositional equivalence and layered geometry of the crystal are confirmed by scanning electron microscopy (SEM, model: Zeiss EVOMA 10) and elemental analysis using energy-dispersive X-ray spectroscopy (EDXS, model: Oxford INCA).

### 2.3 Raman vibrational modes

Raman Spectra was recorded using Renishaw with a Laser of 532 nm (average power below 5 mW). Mechanically cleaved micro-crystal flake was irradiated for 10 sec at a focus of L50X.

### 2.4 Magnetoresistance

A standard four-probe geometry was used to study the transport measurements of $Bi_2Te_3$ and $Sb_2Te_3$ single crystals using the Physical Property Measurement System (PPMS) at various temperatures down to 5 K.

**2.5 Transient reflectance ultrafast spectroscopy (TRUS)**

The TRUS unit consists of a HELIOS spectrometer, Light conversion OPA (optical parametric amplifier), COHERENT amplifier, and MICRA mode-lock laser. The detailed setup is mentioned in the supplementary information.

**3 Results and Discussion:**

**(a)**

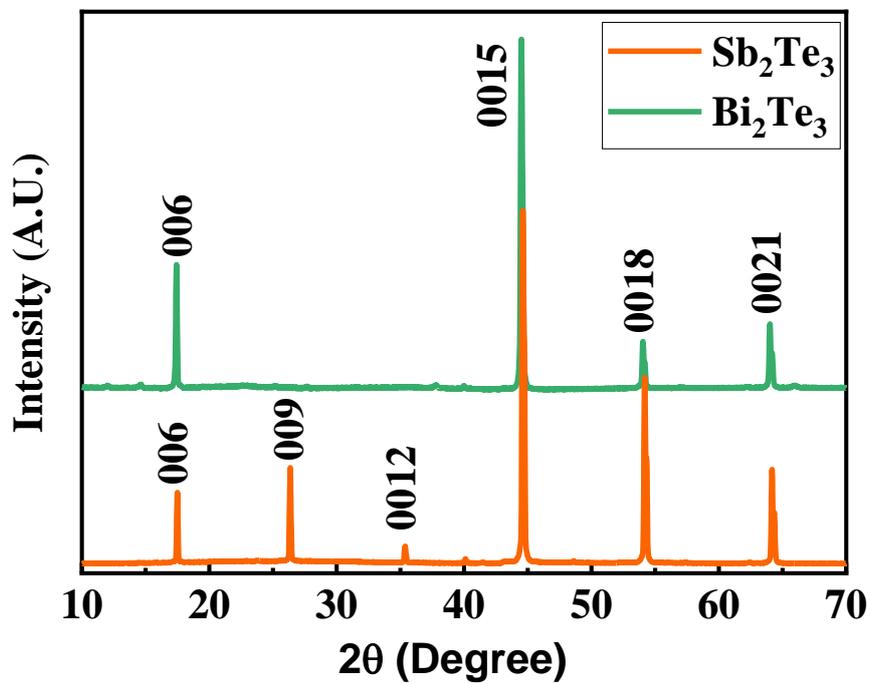

**(b)**

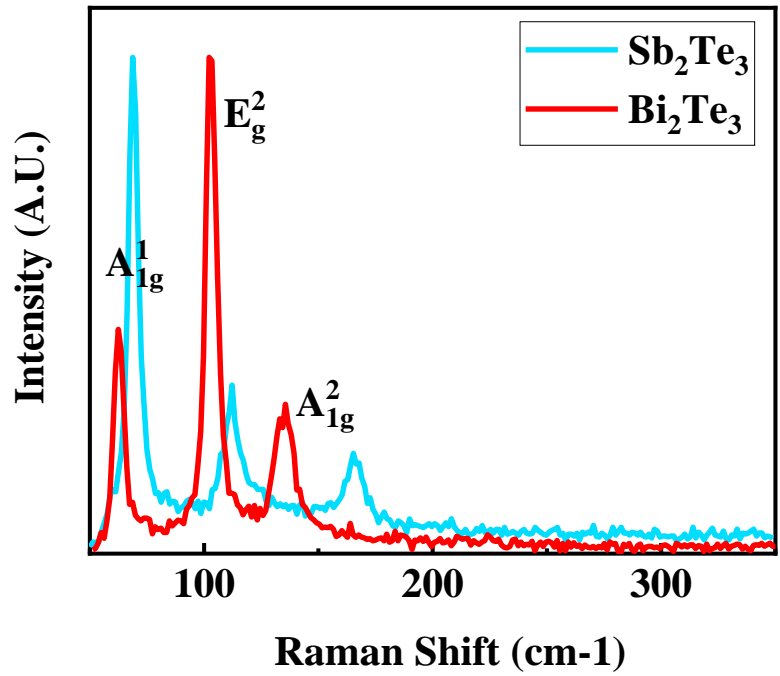

(c)

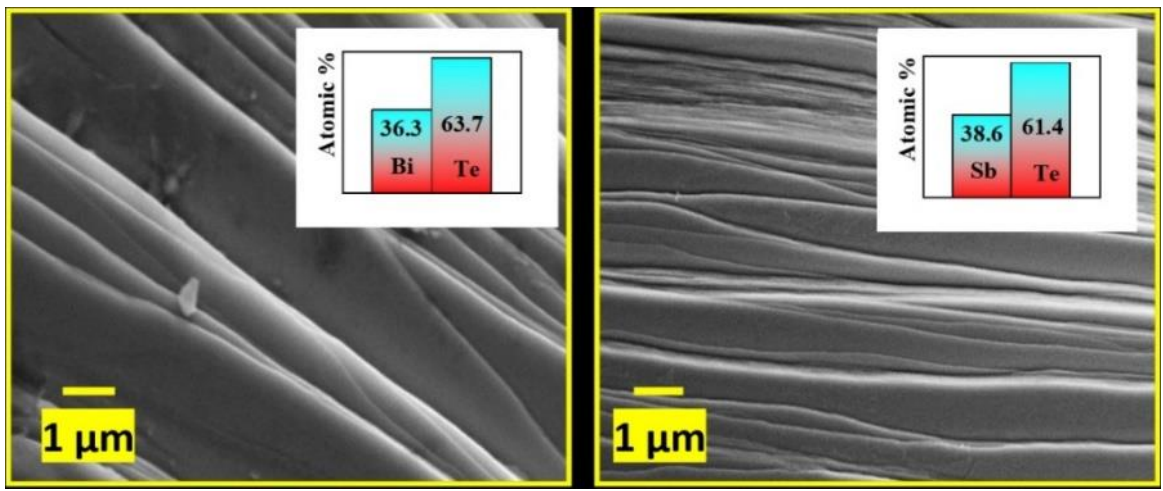

(d)

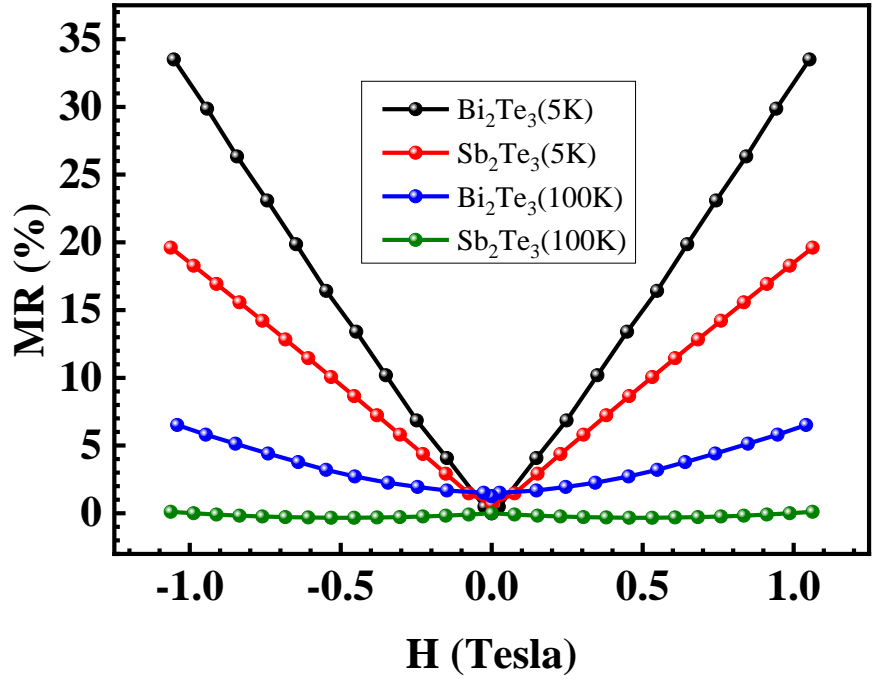

Fig.1 (a) shows the XRD pattern of micro-flakes that are mechanically cleaved through a surgical blade. It shows the (00$l$) planes of both the TIs that confirm the single crystallinity of the grown samples. (b) shows different vibrational Raman modes that are present in these two TIs. (c) shows the SEM images and the EDAX data of both the TIs, confirming the layer-by-layer growth of crystals and their respective composition. (d) shows the percentage magnetoresistance of $Bi_2Te_3$ and $Sb_2Te_3$ single crystals in transverse magnetic field up to 1 Tesla at temperature 5 K and 100 K.

The grown single crystals are mechanically exfoliated through surgical blades to get micro flakes. These flakes are then considered for essential characterization to confirm the structural phase and composition of the TIs. Fig.1(a) shows the XRD pattern of flakes which shows the single-crystalline nature of the TIs. It shows the (00$l$) reflections in both the TIs, which confirms the single crystal growth. The peak positions are also marked in fig.1(a), which ensures the phase formation and structural confirmation of TIs. The vibrational modes are also investigated to confirm the formation, and fig.1(b) shows three different modes of both the TIs. These modes are also in direct accordance with the literature, and hence grown material is found to be bismuth telluride and antimony telluride. The SEM image in fig.1(c) shows the layered structure, which eventually supports the argument of the layered growth of a single crystal. At the same time, the EDAX in the inset of fig.1(c) proposed the compositional equivalence and confirmed the growth of TIs. Coming to the TSS, the magnetoresistance (MR) of $Bi_2Te_3$ and $Sb_2Te_3$ single crystals were measured in a transverse magnetic field of up to 1 Tesla at temperatures 5K and 100K. Fig.1 (d) shows the magnetoresistance of both $Bi_2Te_3$ and $Sb_2Te_3$ crystals at 5 K and 100K.

The percentage magnetoresistance is defined as

$$MR (\%) = [(\rho(H) - \rho(0))/ \rho(0)] \times 100,$$

where $\rho(H)$ and $\rho(0)$ are resistivity at applied and zero magnetic fields, respectively.

Both Bi2Te3 and Sb2Te3 crystals show the positive MR, proportional to the applied magnetic field for measured temperature. At temperature 5 K, the MR reaches ~35% and ~20% for $Bi_2Te_3$ and $Sb_2Te_3$, respectively, under a magnetic field of 1 Tesla. As the temperature is increased to 100K, the percentage MR decreased to ~7% and ~1% for $Bi_2Te_3$ and $Sb_2Te_3$, respectively, at field 1 Tesla. Figure 1(d) shows that at temperature 5K and near-zero magnetic field, the MR shows the v-type cusp behaviour, which corresponds to a weak anti-localization effect due to topological surface states of $Bi_2Te_3$ and $Sb_2Te_3$ crystals[57]. Further, it is observed that with increasing temperature to 100K, the cusp-like behaviour disappears and shows parabolic MR, which suggests that at low temperature, the conduction is dominated by surface states, and with increasing temperature, there is a contribution from both surface states and bulk states in the overall conduction phenomenon[40].

The MR measurements ensure the presence of TSS in these micro flakes, which can be easily seen at the low temperature of 5 K. It can also be concluded from the literature that the bulk states in these flakes are dominant at high temperatures while the surface states are prominent at low temperatures[8]. Considering the same behaviour, the ultrafast charge carrier dynamics of the two TIs are investigated at different temperature in order to study the behaviour of carriers with change in the transformation of the bulk states to TSS with a decrease in temperature from 300 K to 5 K. The micro-flakes are excited with a pump laser of 3.02 eV (410 nm) and probed in NIR regime of 0.77 eV-1.55 eV (1600-800 nm). Fig.2(a) shows the variation of differential reflectance (DR) at different temperatures of 300 K, 200 K, 100 K, 50 K, and 5 K with a probe delay of 500 fs. The DR does not change much with the temperature except for the development of peak at ~0.96 eV (1280 nm) as the TIs are probed below 100 K, which further becomes prominent at 5 K as shown in fig.2(a).

  **(a)**                               **(b)**

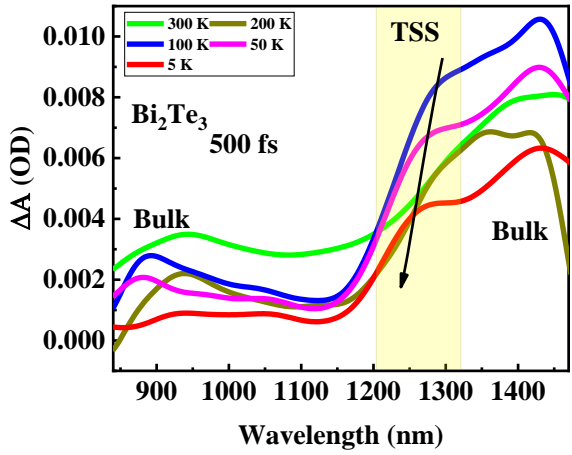
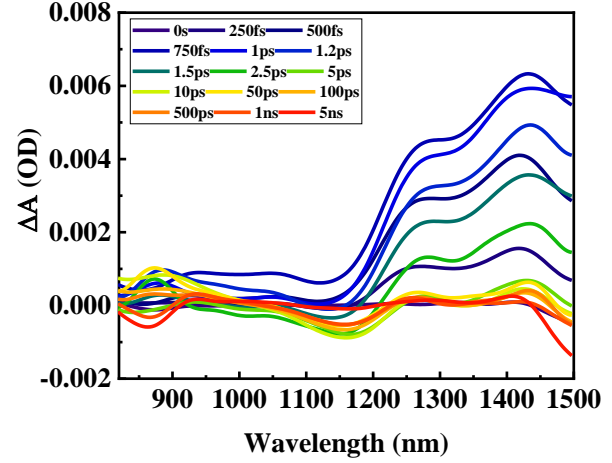

(c)     (d)

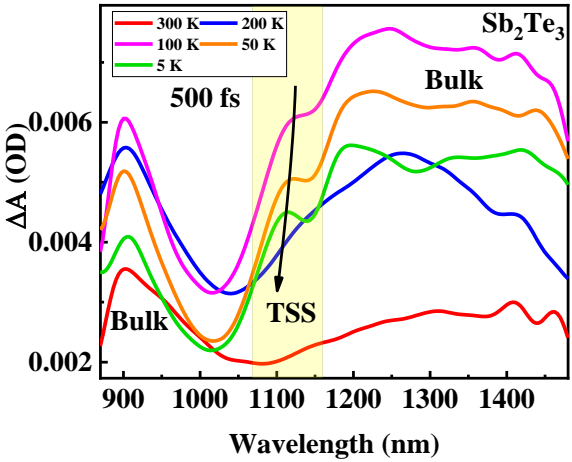
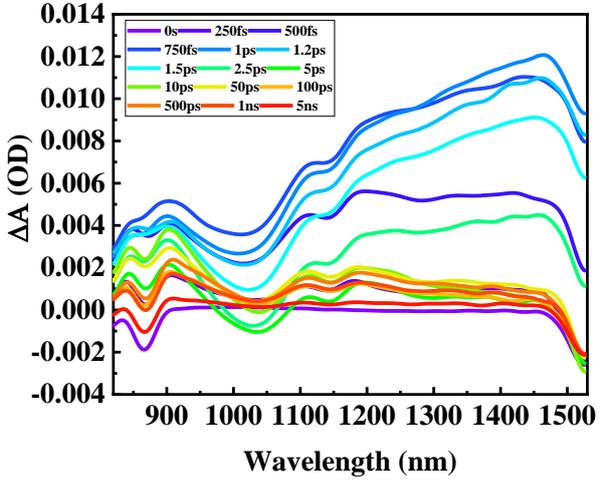

Fig.2 The variation of differential reflectance (DR) concerning temperature at 500 fs is analyzed in which the micro-flakes are excited at 3.02 eV and probed in the 1.55 eV-0.77 eV of different TIs (a) $Bi_2Te_3$ and (c) $Sb_2Te_3$, respectively. The peaks highlighted in the TSS region in both the TIs were found to have blue shift with lowering of temperature which is due to decrease in thermalisation process with the lowering the temperature. At the same time, the probe delay-dependent changes at the same pumped and probe regime but only at 5 K are shown for (b) $Bi_2Te_3$ and (d) $Sb_2Te_3$, respectively.

The temperature-dependent DR shows a peak of ~0.96 eV (~1300 nm) in $Bi_2Te_3$, while a similar peak of ~1.11 eV (~1100 nm) is also observed in $Sb_2Te_3$ as the sample temperature is lowered to 100 K. These peaks become visibly prominent with the further decrease in temperature to 5 K, as shown in fig.2(a) and (c) for $Bi_2Te_3$ and $Sb_2Te_3$ respectively. These emerging peaks at 0.96 eV and 1.11 eV are thus, attributed to the TSS [8] as they appear only at below room temperatures. The MR studies also show that at low temperatures, the TSS dominates over bulk states. However, in TRUS, the spectrum may depict the possibility of defect-dominated states and the states dominated by bulk and TSS. So, to have a closer look at the possibility of these peaks, the probe delay-dependent investigation is analyzed to see the reason for the occurrence of these states and also to trace the origin of other DR peaks that are present in the spectra from room temperature to low temperature 5 K. The DR with probe delay at 5 K is shown in fig.2(b) and (d) for both the TIs, respectively. At the same time, the DR spectra for different temperatures are shown in the supplementary information. Considering the DR spectra of $Bi_2Te_3$ at 5 K in fig.2(b), initially, at 250fs, a broad DR is observed from ~0.7 eV-1.5 eV in which the presence of 1.42 eV, 0.96 eV, and 0.88 eV peaks can be readily noticeable. The magnitude of these peaks increases till 750 fs, resembling the excitation of charge carriers to the excited states that lie at 1.42 eV, 0.96 eV, and 0.88 eV from ground states. It signifies the presence of photoinduced absorption occurring in $Bi_2Te_3$, resulting in many-body effects such as free charge carrier absorption as the TI resembles the metallic nature and thus, has a considerable number of free carriers in the system. After 750 fs, the DR is observed to be decreasing with an oscillating behaviour that is the magnitude of DR sinusoidally increasing and decreasing in the 100 ps range. The decreasing magnitude resembles the relaxation of charge carriers that are excited in initial femtosecond time. While the oscillating component is due to the coherent acoustic phonons (CAP)[20,50,54,58–60]. Near to 1 ps, the negative dip in the DR spectra can be easily noticeable, which resembles the stimulated emission at 1.06 eV, and the relaxation occurs through this stimulated process. The CAP behaviour is also noticeable with the oscillating nature of the stimulated emission peak as well. The CAP is observed in a longer time scale of 10-100 ps, which supports the argument of the occurrence of vibrating acoustic modes [20,58].

While comparing with the excitation of $Sb_2Te_3$, the DR spectra in Fig. 2(d) show broad positive DR magnitude at initial 250fs. The magnitude of DR increases up to 750 fs and eventually decreases. Similar to the $Bi_2Te_3$, the increasing behaviour defines the excitations of charge carriers from ground states to some excited energy levels. As the whole DR is positive, it resembles numerous excitation channels, but the three peaks 1.36 eV, 1.11 eV, and 0.88 eV observed in the DR spectra resemble the prominent channels in particular TI. Again, the positive DR resembles the photoinduced absorption due to many-body effects such as free charge carrier absorption. A negative DR magnitude at ~1.18 eV is observed at 1 ps that support the argument of charge carrier relaxation after 750 fs. Thus, the negative DR has stimulated emission by which the excited carriers are relaxing. The CAP modes are also observed in $Sb_2Te_3$ too. The DR spectra with varying probe delay at 5 K are shown in fig.2(b) and (d) of both the TIs, while the DR spectra at 300K, 200 K, 100 K, and 50 K are presented in supplementary. Comparing the DR spectra at different temperatures as well as at different probe delay, it can be easily seen that the 0.96 eV and 1.11 eV peaks in DR of both $Bi_2Te_3$ and $Sb_2Te_3$ respectively start arising at 100 K and become more prominent at 5 K. Still, the question of this particular transition still not sorted as both the TSS and defects occurring states can also be the reason. However, this can be distinguishable by seeing the carrier's lifetimes. So the kinetics at different probe wavelengths are considered and fitted with particular lifetimes.

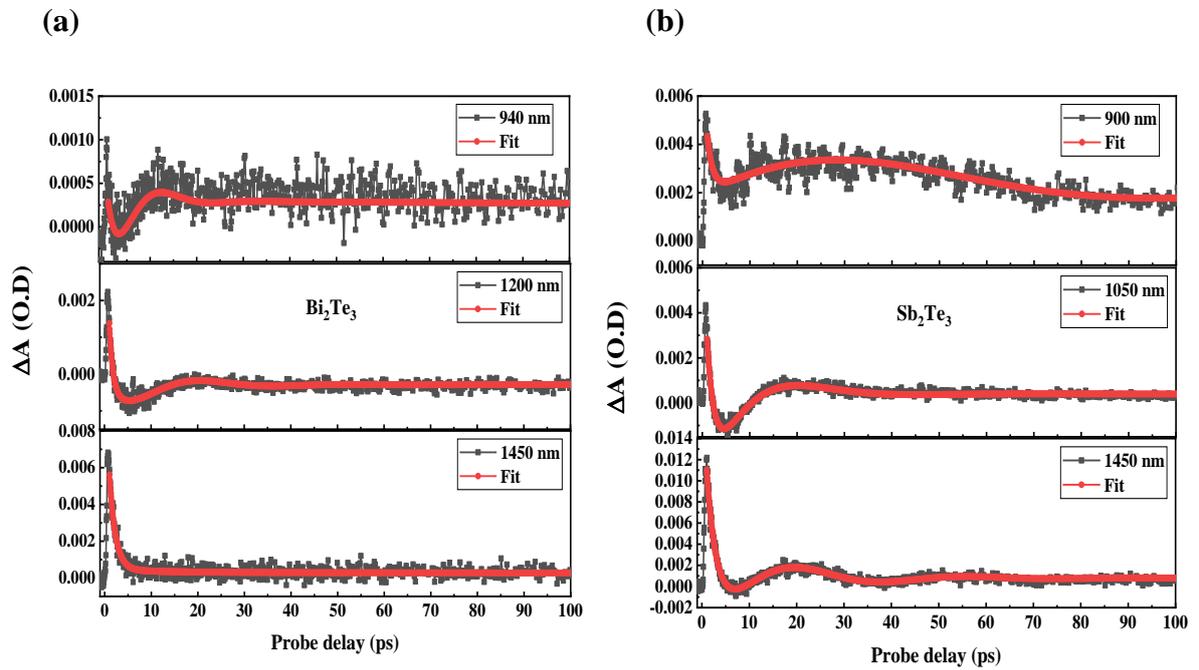

Fig.3. The kinetic profile at three different probe wavelengths of the two TIs (a) $Bi_2Te_3$ and (b) $Sb_2Te_3$ are fitted with two-time constants t1 and t2 along with the sinusoidal damping term that resembles the CAP modes in the two TIs.

The kinetic profile is investigated at three different probe wavelengths and fitted with an equation in which two lifetimes are considered, along with an additional sinusoidal damping term that signifies the acoustic phonon. The fitted equation is

$$C_i^*(t) = a_1 e^{\frac{-t}{t_1}} + a_2 e^{\frac{-t}{t_2}} + b e^{\frac{-t}{c}} cos(2\pi ft + P) + a_0 \qquad (1)$$

Where the cosine term represents the acoustic phonon relaxation, which is a damped harmonic motion as visible in the kinetic profiles, the '$f$' represents the acoustic frequency in the terahertz regime and describes in detail in supplementary information where the t1 is the relaxation lifetime due to interband relaxations, t2 is the final localization of photoexcited carriers and recombination. At the same time, the third -term is a sinusoidal damping term representing the damped acoustics vibrations and relaxation process. The 940 nm (1.31 eV), 1200 nm (1.03 eV), and 1450 nm (0.85 eV) kinetic profiles are considered in $Bi_2Te_3$, and 900 nm (1.37 eV), 1050 nm (1.18 eV), and 1450 nm (0.85 eV) are considered for $Sb_2Te_3$ for fitting profile and determination of the respective lifetimes. Fig.3(a) and (b) show the kinetic profile at these probe wavelengths of $Bi_2Te_3$ and $Sb_2Te_3$ TIs, while the fitting parameters are given in supplementary information. The ease of fitting can be seen through fig—3 (a) and (b), which are shown up to 100 ps. The CAP mode can easily be noticeable in kinetic profiles. Coming to the relaxation lifetimes of $Bi_2Te_3$, the t1 lies in a range less than 1.8 ps. This particular range is close to noble metals[52,54,55,61]. As the sample temperature is reduced from 300 K to 5 K, the t1 also reduces from ~1.8 ps to ~0.8 ps which is in the noble metal carrier relaxation range. This closeness at 5K carrier lifetime resembles the dominant effect of Dirac TSS as the metallic nature is confirmed due to low lifetimes of a few picoseconds, and the only reason for such behaviour found to be existing in $Bi_2Te_3$ is due to its surface states. Similarly, in $Sb_2Te_3$, the t1 is also lying in the same regime, which shows the presence of TSS in the system. While, if trap states or defect states present in the system, then the carrier lifetime t1 does not be so small. The second-lifetime component, 't2', in both the TIs, lies in a few hundreds of ps, which resembles a long relaxation of excited state carriers through recombination.

While in the phonon modes, two types of vibrations are observed in the kinetic profiles. One is the coherent optical phonons (COP) in the 1-10 ps range, and another is coherent acoustic phonons (CAP) in the 10-100 ps range. These modes are also reported in the literature too[20,58]. The vibrational frequency of CAP is found to be lying in the gigahertz range, while the COP lies in the terahertz regime. The CAP frequency is calculated through the kinetic profile fitting in which the frequency of sinusoidal damping resembles the CAP [58,62]. However, the frequency of COP is calculated by subtracting the relaxation kinetics using a high pass filter to find the optical oscillations, and then these oscillations are fitted with sinusoidal damped function as reported in the literature[20,58]. CAP frequency at different temperatures from 300 K to 5 K is illustrated in the gigahertz regime in both the TIs. Such variation is plotted for $Bi_2Te_3$ and $Sb_2Te_3$ as shown in fig.4(a) and (b), respectively. At the same time, the COP at a different temperature from 300 K to 5 K in the Terahertz regime of $Bi_2Te_3$ and $Sb_2Te_3$ respectively is shown in fig.4(c) and (d), respectively. Three different kinetics are considered and examined to determine the carrier relaxations in both the CAP and COP frequency regions. In the $Bi_2Te_3$, a maximum of 55 GHz and a minimum of 1 GHz is observed due to CAP. While for COP, a maximum of 2.0 THz and a minimum of 0.8 THz is observed. While in $Sb_2Te_3$, 50 GHz and 7 GHz frequencies in CAP and 2.5 THz and 1.2 THz frequencies are obtained. The CAP and COP in both the TIs are found to be decreasing with temperature, as shown in Figure 3. The frequency modes in both TIs are also found to be dependent on probe delay, as also reported in $Bi_2Se_3$[20].

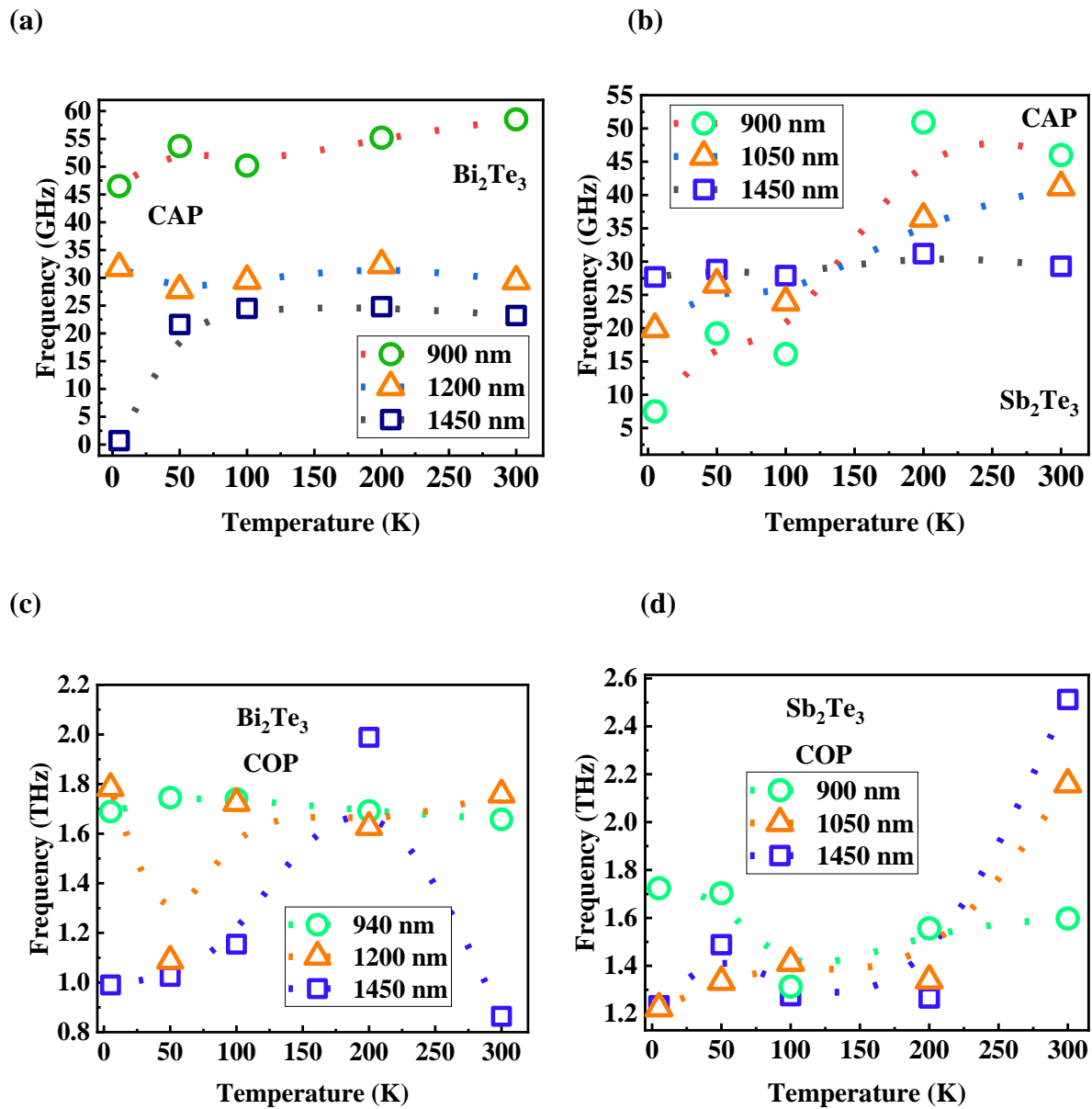

Fig.4. The frequency of CAP at a different temperature from 300 to 5 K that lies in the gigahertz regime of (a) $Bi_2Te_3$ and (b) $Sb_2Te_3$, respectively, are investigated through the kinetic profile and fitting these profiles with carrier lifetime equation. At the same time, the COP at a different temperature from 300 to 5 K that lies in the Terahertz regime of (c) $Bi_2Te_3$

and (d) $Sb_2Te_3$, respectively, is investigated through high pass filtering and damped sinusoidal fitting. These frequencies, whether it is CAP or COP, three different kinetics are considered.

The temperature-dependent DR spectra and kinetic profiles of both the TIs are explored thoroughly, where the various aspects of kinetic profile such as noble metal-related lifetime, CAP, and COP are emphasized. At the same time, the peaks in DR spectra need to be related to the band structure to find the different transitions that are occurring in the two systems along with the temperature-dependent evolution of the peaks at 0.96 eV and 1.11 eV in the respective TIs. Considering the band structure from the literature [63–65] at Γ point and complying with our TRUS results, a model has been proposed where all the transitions occurring in the TIs are manifested.

**(a)**

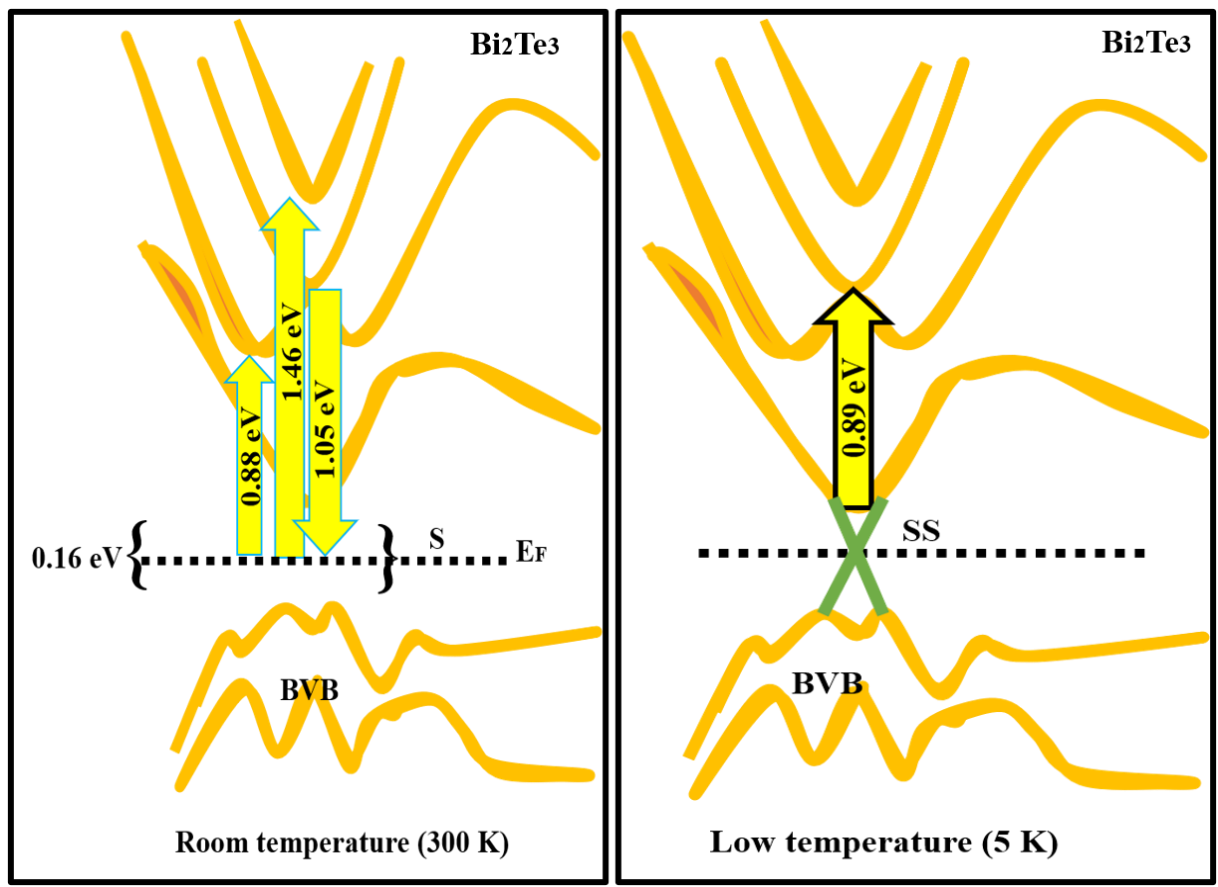

**(b)**

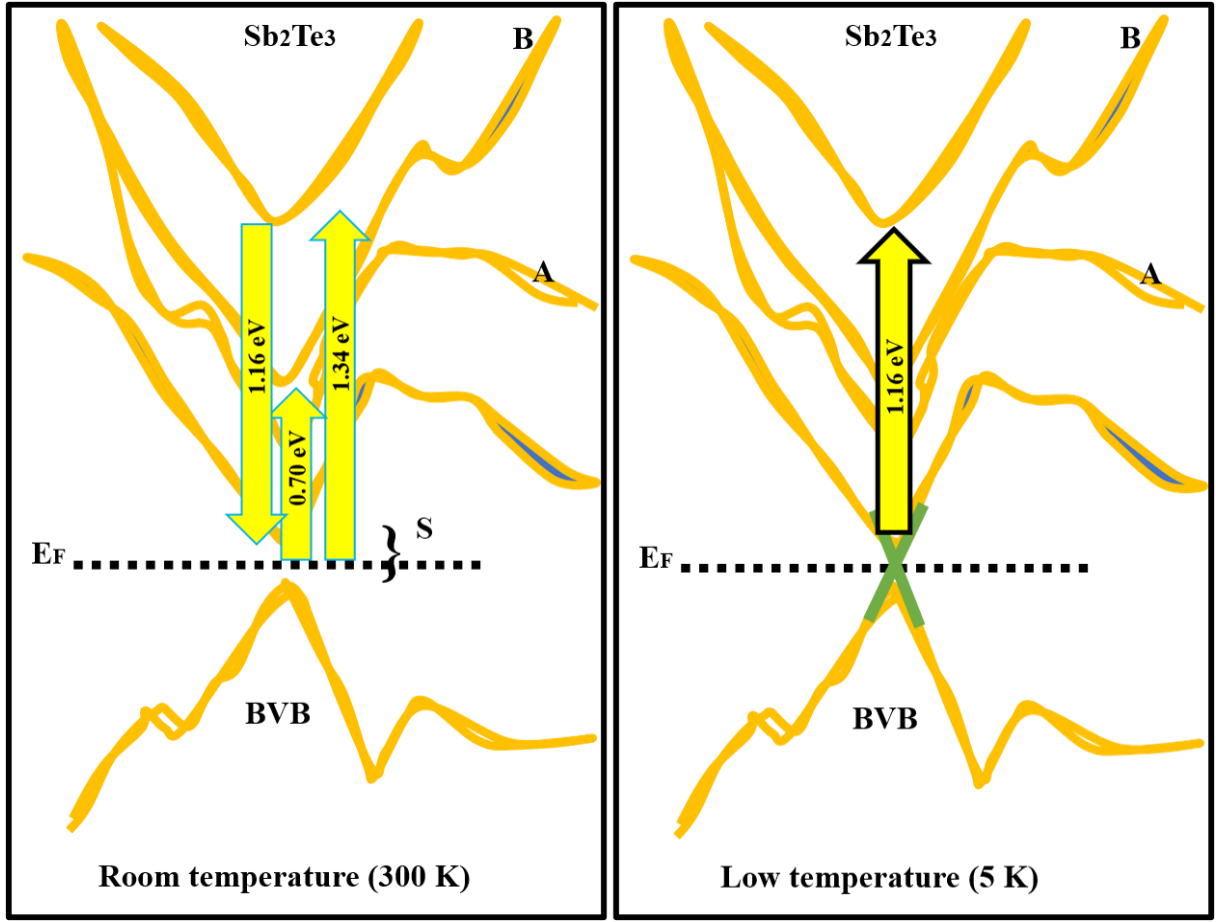

Fig.5 The band near the gamma point is considered in both the TIs where the (a) Bi$_2$Te$_3$ and (b) Sb$_2$Te$_3$. The transitions that are found to be occurring in the TRUS measurements are marked in the schematic diagram.

Considering the Bi$_2$Te$_3$, various optical transitions are observed at in the DR peaks at 0.88 eV, 0.96 eV, 1.06 eV, and 1.42 eV, where it is found that the 0.96 eV peak is observed only at low temperature and is related to dominant surface states at low temperatures. Fig.5(a) shows all these transitions, such that the 1.42 eV transition is the channel where the ground state charge carriers are excited to the bulk conduction band (BCB) due to the pump beam. Similarly, the transition from Fermi level to A band (marked in fig.5(a)) resembles a 0.88 eV DR peak. While the stimulated emission peak of 1.06 eV is a relaxation of excited charge carriers from the B band to the Fermi level. The 0.96 eV, which evolves at a low temperature, depicts the transitions from surface states to the B band. As these 0.96 eV transitions do not appear at room temperature, the charge carriers are still relaxing because of a channel from BCB to B band, which is not visible in our experiment due to experimental limits. While, at low temperature, a metallic channel becomes prominent because of the presence of TSS, and thus, the charge carriers directly move from the lower surface band to the higher surface band. The lifetime also supports the argument of the noble metallic nature of TIs, and thus, the dominance of TSS at low temperatures can be easily seen. Similarly, for the $Sb_2Te_3$, the DR peaks at 0.88eV, 1.11 eV, 1.18 eV, and 1.36 eV are related to the band at Γ point, as shown in Figure 5(b).

**4 Conclusion:**

The micro-flakes of single-crystalline topological insulators $Bi_2Te_3$ and $Sb_2Te_3$, which are grown in-house, are mechanically exfoliated through a surgical blade and investigated for their single crystalline phase, composition, and vibrational modes. The magnetoresistance confirms the occurrence of WAL effect, which depicts the presence of TSS in both samples. At the same time, the presence of prominent surface states is extensively studied and modelled through ultrafast transient absorption spectroscopy. The temperature-dependent phonon modes in both the bismuth telluride and antimony telluride are also established through TRUS. The charge carrier dynamics at 100 K, 50 K and 5 K depict the surface state-related transitions due to increase in surface conduction channel, which are correlated with the low temperature MR studies and demonstrate the low temperature dominant TSS-related transitions.

**5 Acknowledgment**

The director of NPL strongly supports this work. The authors would like to thank Mr. K.M Kandpal for vacuum encapsulation for furnace reactions. The author also thanks CSIR and UGC for financial support and AcSIR for enrolment as a research scholar in its Ph.D. program.

**AUTHOR INFORMATION**

Corresponding Author

Dr. V. P. S. Awana, Email: awana@nplindia.org

Dr. Mahesh Kumar, Email id- *mkumar@nplindia.org*

**Notes**

The authors declare no competing financial interest.